\documentclass[]{spie}  %>>> use for US letter paper
\usepackage{amsmath,amsfonts,amssymb}
\usepackage{graphicx}
\usepackage[colorlinks=true, allcolors=blue]{hyperref}

\usepackage{aas_macros}

\title{Study of Lobster and Kirkpatrick-Baez Designs for a Small Mission dedicated to Gravitational Wave Transient Localization}

\author[a]{John Rankin}
\author[a]{Sergio Campana}
\author[a]{Giovanni Pareschi}
\author[a]{Daniele Spiga}
\author[a]{Stefano Basso}
\author[a]{Marta Maria Civitani}
\author[a]{Paolo Conconi}
\author[a]{Vincenzo Cotroneo}
\affil[a]{Istituto Nazionale di Astrofisica / Osservatorio Astronomico di Brera (INAF/OAB), Via E. Bianchi, 46, 23807, Merate (LC), Italy}

\authorinfo{E-mail: john.rankin@inaf.it}

\begin{document} 
\maketitle

\begin{abstract}
The localization of X-ray counterparts to gravitational wave events requires a telescope with accurate localization capability in a field of view comparable to the region constrained by the gravitational wave detectors. 
In the context of a small, dedicated, mission, we investigate which optical design could satisfy this capability. 
We compare the possible optical designs that have been proposed for X-rays: the Lobster Eye design (both in the Angel and Schmidt variant) --- inspired by the eyes of crustaceans --- consisting of many small capillaries where grazing incidence reflection occurs, the Kirkpatrick-Baez design, where double reflection occurs on two orthogonal parabolic mirrors, and the standard Wolter-I design. We find that the first two designs, compared to the latter, can achieve a significantly larger field of view, and have a good localization capability if the focal length is longer than existing Lobster Eye designs. The Kirkpatrick-Baez design presents the best angular resolution, but the best overall field of view is obtained with a Lobster system: we present a small optical module able to achieve an effective area $>$100~cm$^2$ at 1~keV in a field of view of 10~deg$^2$.

\end{abstract}


\section{Introduction}
The next decades will see a great expansion of gravitational wave detections: the LIGO and Virgo experiment will work at full sensitivity and future gravitational wave interferometers --- such as the Einstein Telescope observatory\cite{2023JCAP...07..068B} --- will come online. Detectable gravitational waves are produced for example in mergers of two compact objects, but a complete view of these events requires to study also the (possible) electromagnetic emission associated. In particular, in the X-rays, it could be possible to localize and observe the event in its initial phases (such as the cocoon afterglow).

Gravitational wave detectors can constrain gravitational wave events to a region of the sky of 10--100~deg$^2$. For an X-ray telescope to rapidly cover such areas with few or no repointings, what is required is a field of view of the order of $\sim$10 deg$^{2}$, while having a constant angular resolution of at least $\sim$50--100 arcsec over the entire field of view --- so allowing to accurately identify the source position. Table \ref{tab:mission_reqs} reports the mission requirements that we consider as the baseline in the remaining of this work. 

\begin{table}
\caption{\label{tab:mission_reqs} Mission requirements, imposed in the remaining of this work, for a small mission dedicated to accurately localizing gravitational wave events electromagnetic counterparts.}
    \begin{center}       
    \begin{tabular}{c|c}
    Field of view&  $>$10~deg$^2$ = 1.8~deg radius \\ \hline
         Effective Area &  $>$100~cm$^2$  across the field \\ \hline
         Half Energy Width&  $<$50-100~arcsec across the field 
    \end{tabular}
    \end{center}
\end{table}

There is no current X-ray telescope satisfying the scientific requirements of Table \ref{tab:mission_reqs}. Furthermore, the current optical designs would require a substantial combination of different telescopes to achieve these requirements. In this paper we investigate instead which optical design could be appropriate for a small mission (for example ESA Fast (F) class or NASA small explorer (SMEX)) dedicated to accurate gravitational wave counterpart localization in the X-rays.

The last half century has seen a great use of the Wolter-I grazing-incidence optical design, based on double reflections on a parabolic and then hyperbolic surface; most X-ray missions have used it. 
This design can be optimized for wide fields with little off-axis degradation of the angular resolution by expanding the surface equations as polynomials and fine-tuning their parameters. While these polynomial optics have been studied at INAF-Merate Observatory\cite{2010MNRAS.405..877C} for achieving wide fields, 
this requires short mirror lengths so far, which are challenging to manufacture. These studies were also limited to $\sim$1~deg$^2$ field of view (smaller than what is required for the case under study in this work), even if the concept is probably extendable to larger field of view at the cost of an important technological effort.
In this paper, we compare alternative optical designs, focusing on two alternatives which are potentially cheaper than the standard Wolter-I design.
The first option we present is based on the Lobster Eye\cite{1979ApJ...233..364A, 1975NucIM.127..285S} --- named for its resemblance to the eyes of crustaceans. 
We model such a telescope with a longer focal length to aperture ratio than usual, so to obtain a large field of view (suitable for accurate transient localization) but smaller with respect to previous Lobster Eye designs, suited as all-sky monitors. We also present the alternative Kirkpatrick-Baez design\cite{1948JOSA...38..766K}. We then compare these designs with the standard Wolter-I design.

This work is structured as follows. We describe the Lobster Eye design in Section
\ref{sec:Lobster-Eye-Design}, where we discuss the different ways in which it can
be arranged and optimized, and then we describe the Kirkpatrick-Baez design in Section \ref{sec:Kirkpatrick-Baez-Design}.
We compare our design options between themselves and with the standard Wolter-I design in
Section \ref{sec:Comparison-with-Other}, and in Section \ref{sec:Conclusion} we draw our conclusions.

\section{Lobster Eye Designs}\label{sec:Lobster-Eye-Design}

The Lobster Eye concept\cite{2021hai4.book...85W} is based on small square channels which focus photons through gracing incidence double reflections on two flat orthogonal surfaces. Two configurations have been proposed inspired by Lobster Eyes for X-ray astronomy: the Angel design\cite{1979ApJ...233..364A} and the Schmidt design\cite{1975NucIM.127..285S}

\begin{figure*}
\begin{center}
\includegraphics[width=0.39\linewidth]{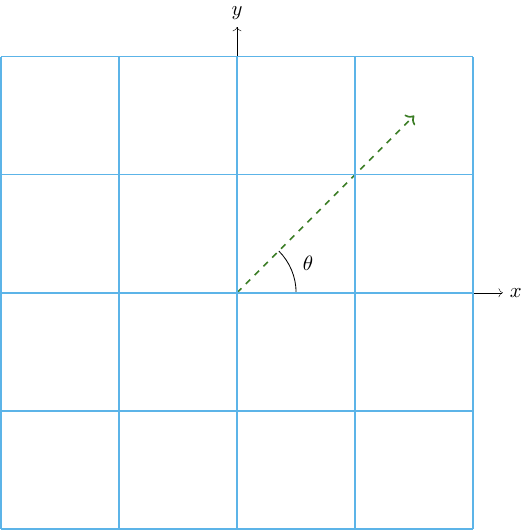}
\includegraphics[width=0.59\linewidth]{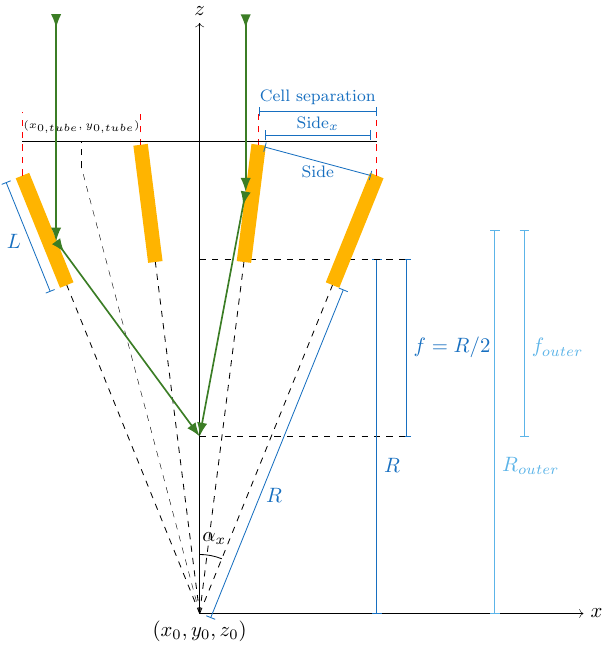}
\end{center}
\caption{\label{fig:Schematic-of-Lobster-this-paper} Schematic (not to scale) of a portion of a Lobster Eye mirror module and conventions used in this work. Left:
top view; right: side view. In the Schmidt case, the $R$ and $f$ represented correspond to $R_{inner}$ and $f_{inner}$, while the outer curvature and focal point are shown in light blue. On the right only a section is shown, so only one of the two orthogonal reflections is represented.
 }
\end{figure*}

The Angel design consists of an array of small square capillary tubes with reflecting
surfaces distributed on a spherical surface; if their side is much
longer than their width, reflections at grazing incidence occur, making
this system ideal for X-rays. The reflected rays focus on a sphere
of half-radius of the sphere of the capillary tubes surface. Figure \ref{fig:Schematic-of-Lobster-this-paper}
shows a 1-dimensional schematic of a Lobster Eye.
Thanks to the spherical geometry, the field of view can be increased
just by taking a larger portion of a sphere (i.e. more capillary tubes), while the angular resolution
can be improved by increasing the curvature radius and by decreasing
the size of the capillary tubes. 

The Schmidt design has very similar performance, but the
orthogonal reflecting surfaces are separated into two arrays of parallel
mirrors, potentially making the fabrication easier.

We will now describe the equations to model and simulate a Lobster Eye. All simulations in this work are performed by geometrical optics ray tracing: we describe rays as lines and the interaction of X-rays with surfaces using Fresnel and Snell's equations, computing the reflectivity considering the surfaces to be made of gold. We start with a flat surface, described as
\begin{equation}
\begin{cases}
x=\tan\left(\alpha\right)\left(z-z_{0}\right)\cos\left(\theta\right)+\sin\left(\theta\right)t+x_{0}\\
y=\tan\left(\alpha\right)\left(z-z_{0}\right)\sin\left(\theta\right)+\cos\left(\theta\right)t+y_{0}
\end{cases}
\end{equation}
where $\alpha$ and $\theta$ are, respectively, the inclination and
azimuthal angle of the flat surface, and $t$ is a parametric variable
used to plot the surface; the surface is so designed to pass through
the $\left(x_{0},y_{0},z_{0}\right)$ point.

Given a grid of points (the centers of each cell, see Figure \ref{fig:Schematic-of-Lobster-this-paper}
left), considering the spherical surface we have
\begin{equation}
\alpha=\arctan\left(\frac{\sqrt{x_{0\;tube}^{2}+y_{0\;tube}^{2}}}{R}\right)
\end{equation}
\begin{equation}
\theta=\arctan\left(\frac{x_{0,tube}}{y_{0,tube}}\right)
\end{equation}
where $x_{0,tube}$ and $y_{0,tube}$ are the coordinates of points,
at vertical coordinate $z=R$, through which lines perpendicular to
the spherical surface of the Lobster pass, and $R$ is the curvature
radius of the Lobster surface. The inclinations over the two axis
are then (Figure \ref{fig:Schematic-of-Lobster-this-paper} right)
\begin{equation}
\begin{cases}
\alpha_{x}=\arctan\left(\tan\left(\alpha\right)\cos\left(\theta\right)\right)\\
\alpha_{y}=\arctan\left(\tan\left(\alpha\right)\sin\left(\theta\right)\right)
\end{cases}
\end{equation}
the sides of each cell (in the Angel case) are then given by 
\begin{equation}
\begin{cases}
S_{x}=\frac{S}{\cos\left(\alpha_{x}\right)}\\
S_{y}=\frac{S}{\cos\left(\alpha_{y}\right)}
\end{cases}
\end{equation}
For the Schmidt case, the cells are not square, but one
side in each layer is long as the entire telescope, while the other side is given by this equation.

The walls are positioned so that their planes pass through the coordinates
$x_{0}\pm\frac{S_{x}}{2}$ and $y_{0}\pm\frac{S_{y}}{2}$ depending
on their orientation. The walls are then defined by specifying their
borders in three dimensions; the vertical $\left(z\right)$ borders
are defined as 
\begin{equation}
R_{curvature}\cos\left(\alpha\right)-z_{offset}<z<\left(R_{curvature}+L\right)\cos\left(\alpha\right)-z_{offset}
\end{equation}
where $R$ is the radial distance from the center, $L$ is the length
of each tube, and 
\begin{equation}
z_{offset}=R_{curvature}-R
\end{equation}
is an offset: for the Angel case $R_{curvature}=R$; for the Schmidt case there are two orthogonal arrays, with different radii of the upper and lower layers: $R_{outer}$ and $R_{inner}$. For the bottom Schmidt array we still have $R_{curvature}=R_{bottom}$, but for the top Schmidt array the radial distance from the center reference is not the same as the curvature radius of the spherical Lobster surface, in this case (see Figure \ref{fig:Schematic-of-Lobster-this-paper})
\begin{equation}
R_{curvature}=2f_{outer} = 2\left(R_{outer}-\frac{R_{inner}}{2}\right)
\end{equation}
(basically we are computing two times the distance to the focus, which is set to be the same for both the outer and inner array, and we are considering that the focal point is $R_{inner}/2$ distant from the origin).

In the $x$ and $y$ plane the boundaries considered are
\begin{equation}
x_{0}-\frac{S_{x}}{2}+z\tan\left(\alpha_{x}\right)<x(z)<x_{0}+\frac{S_{x}}{2}+z\tan\left(\alpha_{x}\right)
\end{equation}
\begin{equation}
y_{0}-\frac{S_{y}}{2}+z\tan\left(\alpha_{y}\right)<y(z)<y_{0}+\frac{S_{y}}{2}+z\tan\left(\alpha_{y}\right)
\end{equation}

The baseline parameters we use in this work are reported in Table \ref{tab:Optical-parameters-used}

\begin{table*}
\caption{\label{tab:Optical-parameters-used}
Optical parameters used in this work for the different designs. }
\begin{center}
\begin{tabular}{c|ccccc}
\multicolumn{1}{c}{} & \multicolumn{1}{c|}{Angel} & \multicolumn{1}{c|}{Schmidt} & \multicolumn{1}{c|}{Kirkpatrick-Baez} & Wolter-I\\
\hline 
Curvature radius  & \multicolumn{1}{c|}{5~m} & \multicolumn{1}{c|}{5~m, 4.9~m} & \multicolumn{1}{c|}{N.A.} & N.A.\\
\hline 
Separation of the two layers & \multicolumn{1}{c|}{N.A.} & \multicolumn{1}{c|}{0.1~m} & \multicolumn{1}{c|}{0.1~m} & N.A.\\
\hline 
Focal distance &  \multicolumn{4}{c}{2.5~m}  \\
\hline 
Energy & \multicolumn{4}{c}{ 1~keV}  \\
\hline 
Mirror length & \multicolumn{1}{c|}{ 40~mm } & \multicolumn{3}{c}{ 40~mm + 40~mm }  \\
\hline 
N mirrors &   \multicolumn{3}{c|}{ 250 per side } & 138 shells  \\
\hline 
Mirror separation & \multicolumn{4}{c}{ 0.75~mm  }  \\

\hline 
Mirror thickness &  \multicolumn{4}{c}{ 0.35~mm  }  \\
\hline 
Mirror material & \multicolumn{4}{c}{ Au }  \\
\hline 
Geometric area & \multicolumn{3}{c|}{27.5$\times$27.5~cm$^{2}$} & 13.8~cm radius\\
\end{tabular}
\end{center}
\end{table*}

\begin{figure}
\begin{center}
\includegraphics[width=0.49\linewidth]{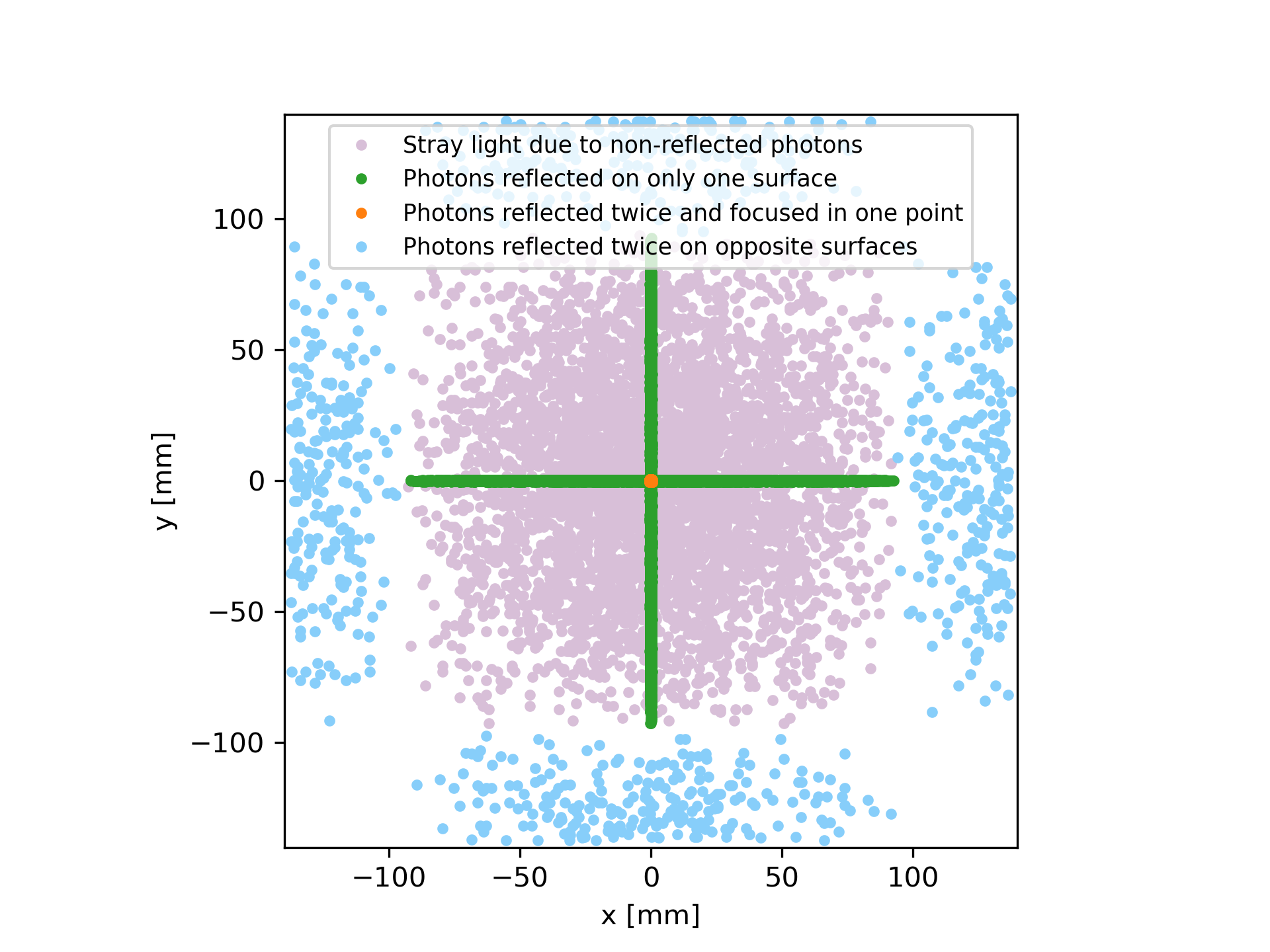}
\includegraphics[width=0.49\linewidth]{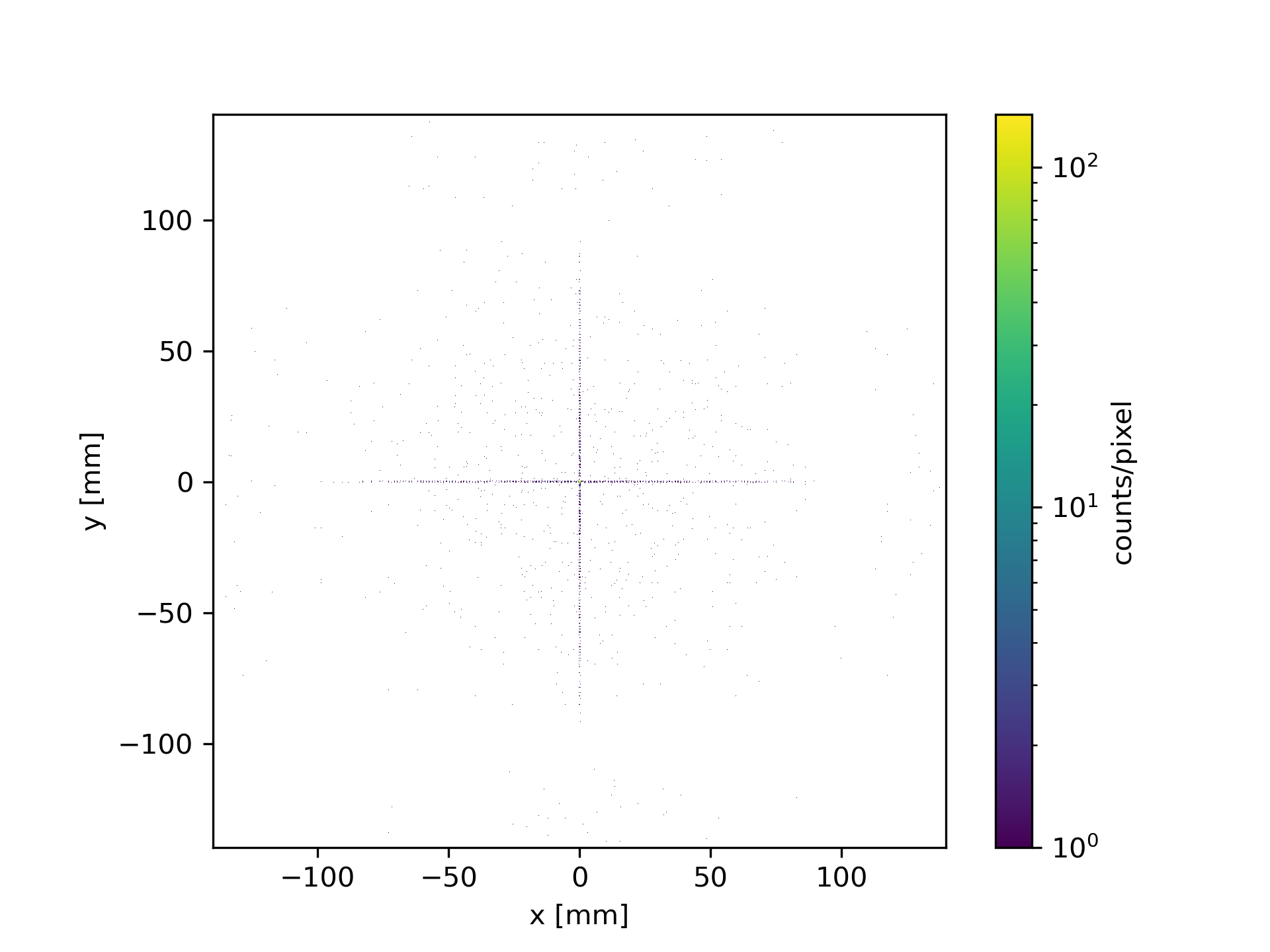}
\end{center}
\caption{\label{fig:Color-coded-simulation} 
Distribution of photons in the image formed from a Lobster Eye. Left: the different colors indicate the number of reflections of photons (but not the density per pixel); the fractional distributions of photons depends
on the optical configuration (see for example Figures \ref{fig:L_optimization} and \ref{fig:Lobsters_comparison}
below). Right: pixel-binned map of the image, showing the density per pixel of the different components.
}
\end{figure}

The formation of the image is different from that of conventional optics, leading to a unique point spread function characterized by a central focus and distinct ``arms''.
Rays can undergo various possibilities:
\begin{itemize}
\item Not hit any surface (stray light).
\item Reflect only one surface: in this case the rays will focus only in
one of the two dimensions, forming two arms of a cross centered at
the focus.
\item Reflected twice on two orthogonal surfaces:
the rays are focused in the center, in a square with size equal to
the size of the square tube divided by the focal distance.
\item Reflected twice on two opposite surfaces (note that all mirrors in Lobster designs are normally double reflecting).
\end{itemize}
A simulation color coded with these different possibilities is shown in Figure \ref{fig:Color-coded-simulation}. The effective area, defined as the product of the geometrical area of the system and the fraction of photons arriving on the detector (i.e. telescope efficiency) is computed using all photons that focus in the center and in the cross (which also contains information on the localization). We also compute the half energy width, usually defined as the radius of the circle containing half the impinging photons, using the thickness of the cross in the $x$ and $y$ projections (so taking into account the peculiar shape of the point spread function of these optical systems).

The photons reflected twice on two opposite surfaces form  ``clouds'' visible in the left side of this figure. The fraction of these photons over the total of the photons that arrive to the detector are reported as ``fraction\_clouds'' in Figures  \ref{fig:L_optimization} and \ref{fig:Lobsters_comparison}; this fraction is consistently negligible. 
Nevertheless it is possible to reduce or avoid this effect in different ways:
\begin{itemize}
\item By having only one side of the mirrors reflecting (e.g. having a non-reflecting outer side with respect to the center of the telescope array). We found however that this reduces the effective area.
\item By having the length of the cells go as $L=S/\tan(\alpha)$; but this requires the cell lengths to become inconveniently long.
\item By having the size of the cells go as $S=L\tan(\alpha)$; but this causes a worsening of the angular resolution due to the wide outer cells.
\end{itemize}
For these reasons we consider more convenient to keep this small additional source of spurious photons present (it is basically an addition to the distributed stray light cause by photons that do not undergo reflections).

Some Lobster Eye telescopes have already been used in X-ray astronomy: these
on-board the Lobster-Eye X-ray Satellite\cite{2022ApJ...941L...2Z}
and on-board the Einstein Probe mission\cite{2022hxga.book...86Y}
have a wide field of view (3600~deg$^{2}$ for the Einstein Probe),
but modest angular resolution (5 arcmin for the Einstein probe) and small effective area ($\sim$2~cm$^2$ at 1~keV), 
so are suitable as all-sky monitors. Also the SVOM mission (Space-based multi-band
astronomical Variable Objects Monitor) has on board a Lobster Eye telescope
--- the MXT telescope (Microchannel X-ray Telescope\cite{2023ExA....55..487G})
--- with a small field of view ($58\times58$~arcmin$^{2}$), but
also a small focal length, and thus modest angular resolution ($\lesssim$
2~arcmin).

The Lobster Eye of this work has instead a longer focal length (2.5~m), and so a more modest (compared to Einstein Probe) field of view of  $\sim$10~deg$^{2}$ but with improved angular resolution and effective area
--- suited as an accurate transient localizer. Table \ref{tab:Comparison-of-our}
compares our design with the different existing ones.

\begin{table*}
\begin{center}
\caption{\label{tab:Comparison-of-our} Comparison of our design with existing Lobster Eyes. Our design has a longer focal length and improved angular resolution and effective area compared to the others, but the field of view is more modest compared to Einstein Probe.}
\begin{tabular}{c|c|c|c}
 & Our design & Einstein probe & SVOM/MXT\\
\hline 
Focal distance & 2.5~m & 0.375~m & 1.15~m\\ \hline 
Field of view & 10~deg$^{2}$ & 3600~deg$^{2}$ & 1~deg$^{2}$\\ \hline 
Effective Area & 100--200~cm$^{2}$ & 2--3~cm$^{2}$ & 35~cm$^{2}$\\ \hline 
Angular resolution & 0.5--1~arcmin & 5~arcmin & $<$ 2 arcmin
\end{tabular}
\end{center}
\end{table*}

\subsection{Optimization of a Lobster Angel Design \label{subsec:Angel_optimize}}

In this section we optimize the Lobster Eye design so that it is appropriate for the objective presented in the introduction, in particular having a field of view of 10~deg$^2$, and we also constrain the effective area $>$100~cm$^2$. The same optimization results have also been found for the similar case of a Lobster-Schmidt, and so are not reported here for conciseness.

\begin{figure}
    \begin{center}
    \includegraphics[width=0.65\linewidth]{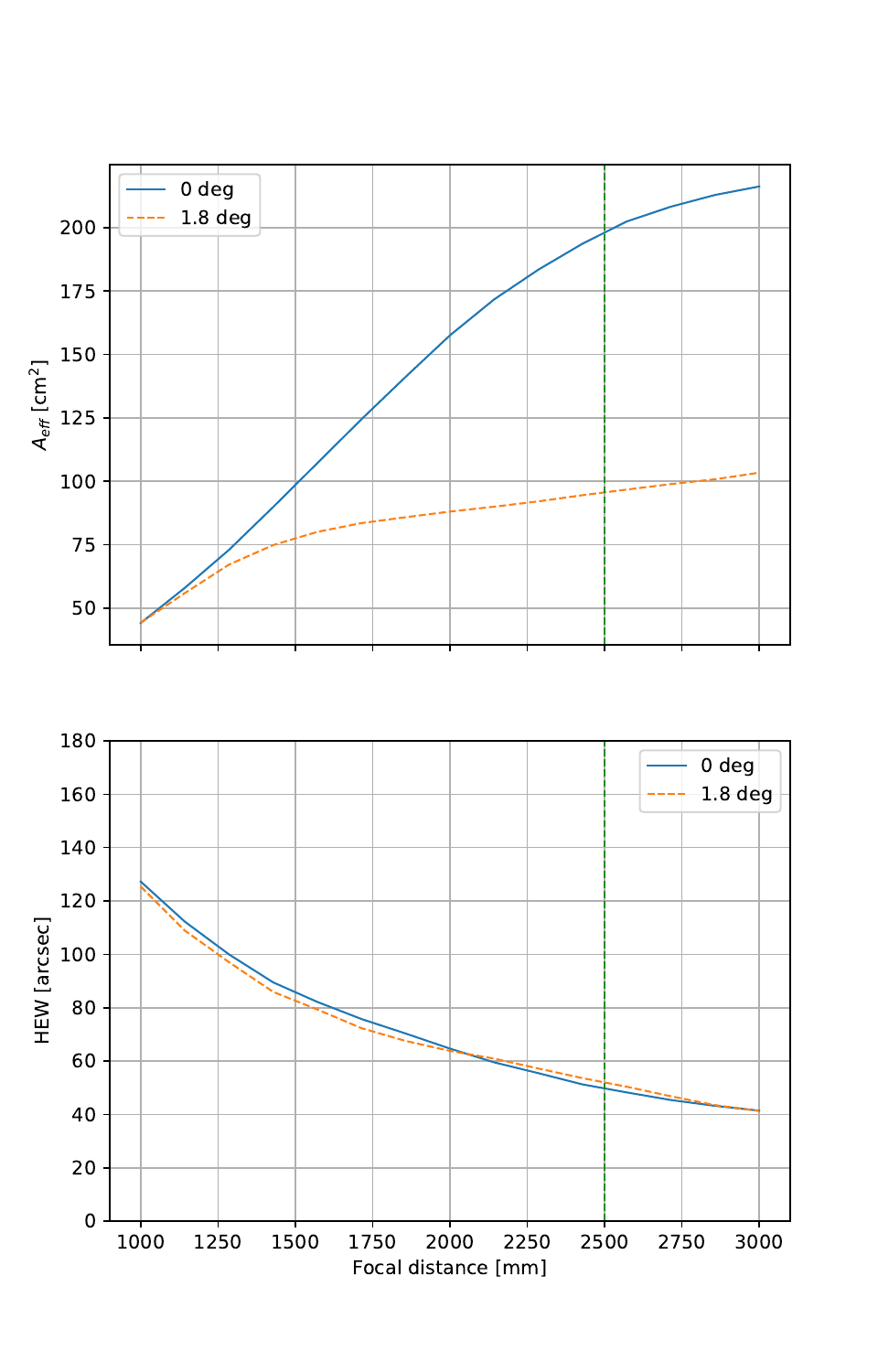}
    \end{center}
    \caption{\label{fig:f_variations} Effective area and half energy width in a Lobster Angel telescope, for different focal distances. The continuous blue curve refers to an on-axis source, the dashed orange curve to a source 1.8~deg off-axis.
     The optical parameters are reported in Table \ref{tab:Optical-parameters-used}; the green vertical lines indicate the value used in the remaining of this work.
     }
\end{figure}

\paragraph{Focus Optimization}
In Figure \ref{fig:f_variations} we show the effective area and half energy width, as a function of off-axis angle, for different focal lengths. We can see that the angular resolution strongly improves with longer focal lengths, and some improvement is visible also for the effective area, particularly off-axis, due to the smaller grazing incidence angles. Satellite payload length constrains will also limit a practical implementation. In the remaining of this paper we consider a focal length of 2500~cm (corresponding to a curvature radius of 5000~cm).

\paragraph{Variation of Mirror Length}

\begin{figure}
\begin{center}
\includegraphics[width=0.65\linewidth]{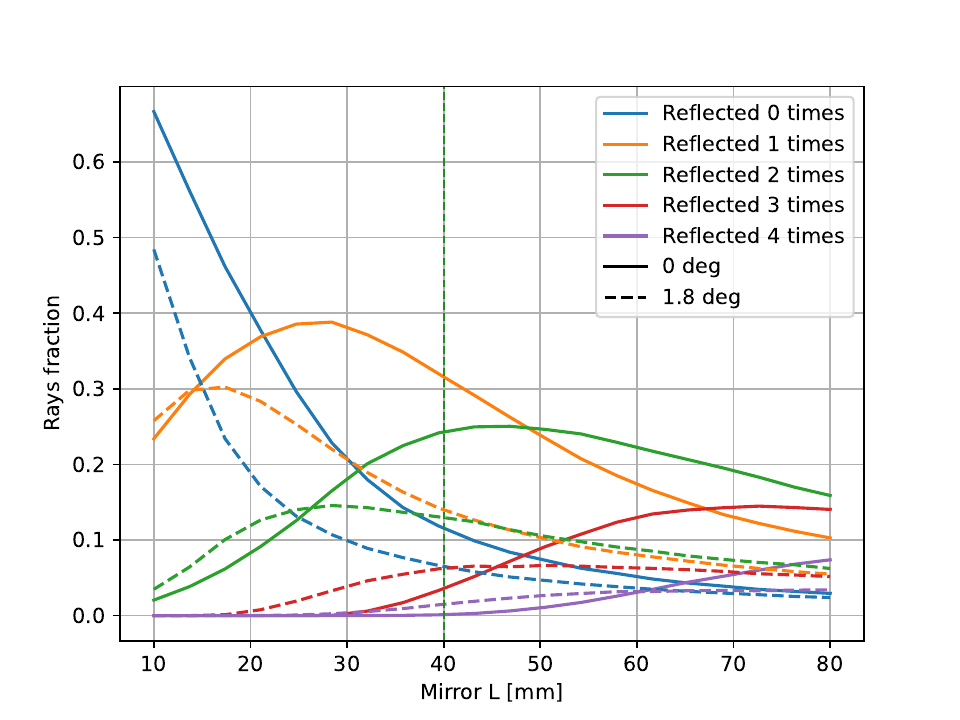}
\end{center}
\caption{\label{fig:Number-of-rays} Fraction of rays reflected on-axis for each reflection order, for the
Angel case, as a function of mirror length.  The continuous curves refer to an on-axis source, the dashed curves to a source 1.8~deg off-axis. For the remaining of this work we chose a length of 40~mm (indicated by the the green vertical lines).
The optical parameters are reported in Table \ref{tab:Optical-parameters-used}.}
\end{figure}

\begin{figure}
\begin{center}
\includegraphics[width=1\linewidth]{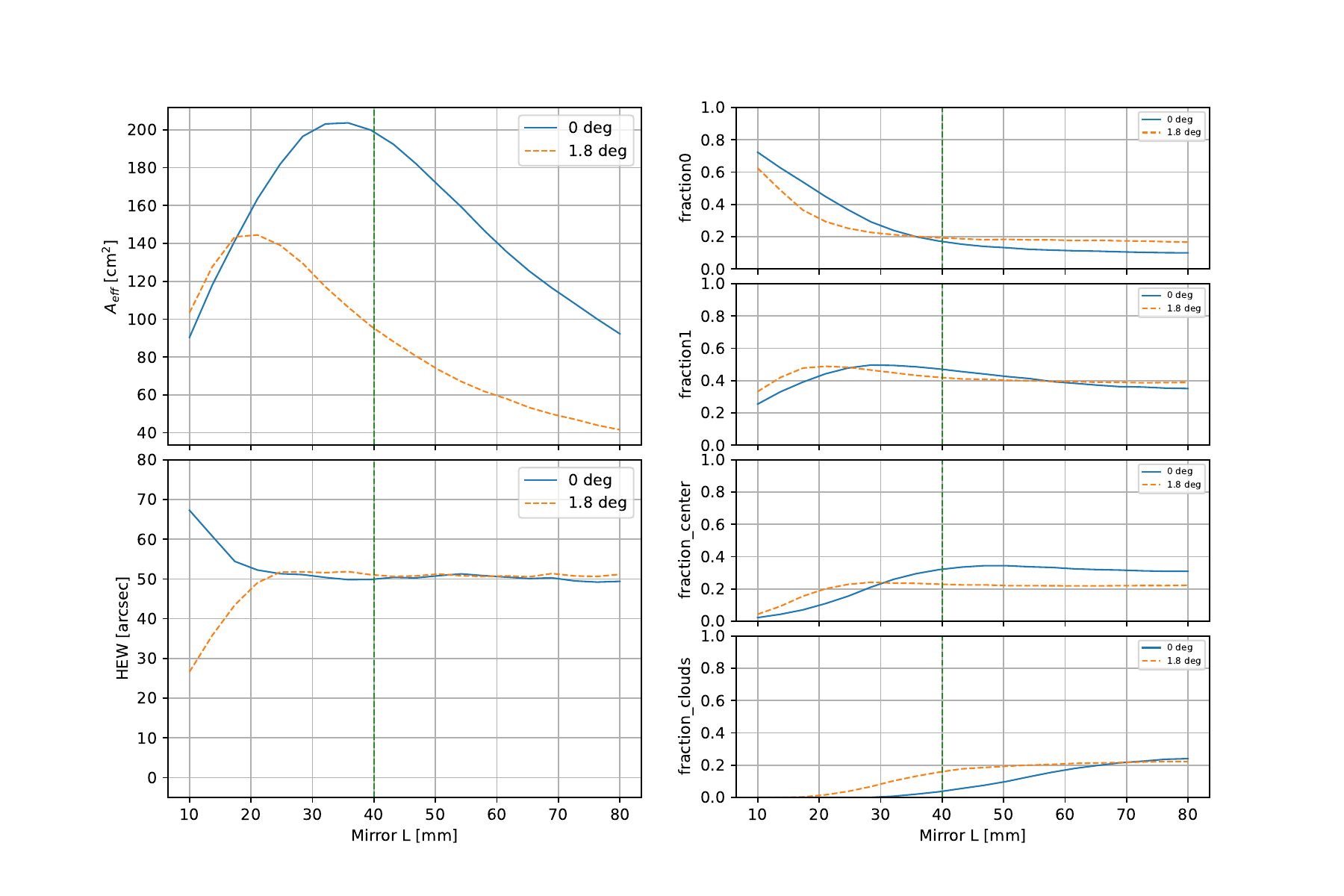}
\end{center}
\caption{\label{fig:L_optimization} 
Performance in effective area, half energy widths and fractions of rays, for the Angel case, as a function of mirror length. The continuous blue curve refers to an on-axis source, the dashed orange curve to a source 1.8~deg off-axis. The fractions of rays refer to, from top to bottom, rays that do not reflect on any surface (stray light), rays reflected once (forming the arms of the cross), rays double reflected to the focusing center and rays double reflected outside of the focusing center (in the ``clouds'', see Figure \ref{fig:Color-coded-simulation}).
For the remaining of this work we chose a length of 40~mm (indicated by the the green vertical lines).
The optical parameters are reported in Table \ref{tab:Optical-parameters-used}. }
\end{figure}

Figure \ref{fig:Number-of-rays} shows how many rays are reflected on-axis for each reflection order as a function of the length of mirrors, while Figure \ref{fig:L_optimization} shows the performance: we see that the effective area on-axis peaks at $L=$40~mm, while off-axis the peak is at smaller lengths. However we also see that the stray light and fraction of useful photons are better at lengths $L>$40~mm. Based on this, the simulations in this work were performed with a mirror length of 40~mm, which maximizes the fraction of double reflected rays.

\begin{figure}
\begin{center}
\includegraphics[width=0.65\linewidth]{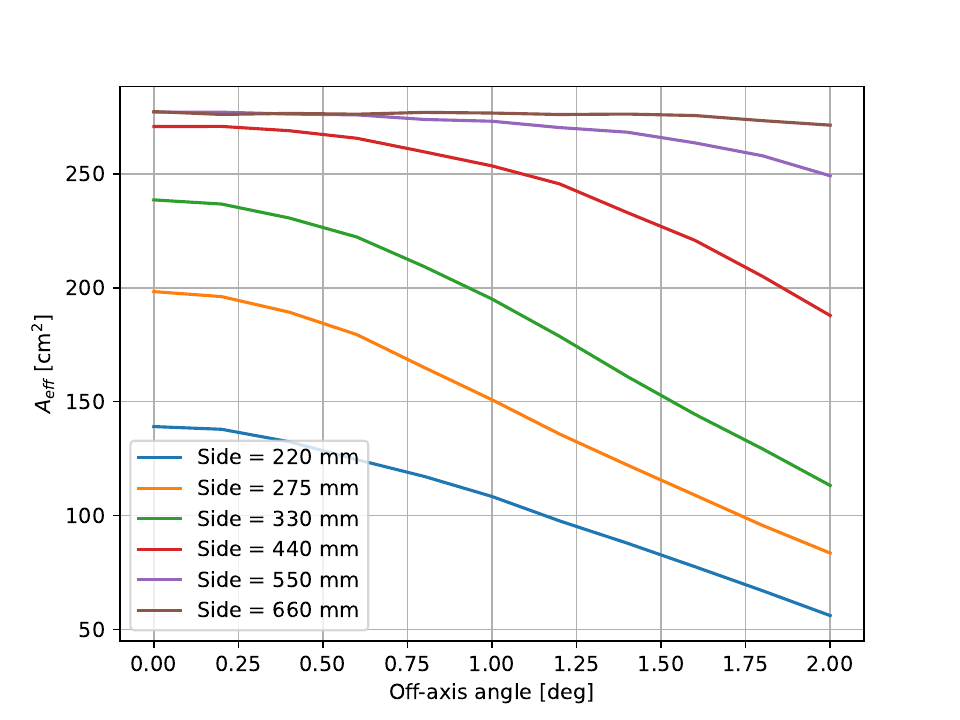}
\end{center}
\caption{\label{fig:Lobster_max_size}
Performance in effective area for the
Angel case, for different physical sizes (given by the cell distance multiplied by the number of cells per side) of the system: increasing this variable improves the vignetting more than the on-axis effective area. The optical parameters are reported in Table \ref{tab:Optical-parameters-used}. }
\end{figure}

\paragraph{Variation of telescope size}
Figure \ref{fig:Lobster_max_size} shows the effective area of a Lobster telescope as a function of its physical size (given by the cell distance multiplied by the number of cells per side): as the size increases, the effective area on-axis increases only up to a point, while the biggest improvement is in the reduction of vignetting. This is because the off-axis cells have a limited contribution to the on-axis effective area, which depends then mainly on the radius of curvature (see also Figure \ref{fig:f_variations}).

\section{Kirkpatrick-Baez Design}\label{sec:Kirkpatrick-Baez-Design}

Another optical design, alternative to Wolter-I or Lobster, is the Kirkpatrick-Baez design\cite{1948JOSA...38..766K},
which consists of two arrays of parabolic (on one axis) reflecting
surfaces: the first array reflects
the X-rays on a line, which are then focused to a point by the second
--- orthogonal --- array (this design is basically a variation of
the Schmidt Lobster Eye with parabolic surfaces).

The parabola in the $z-x$ reference, where $z$ is the optical axis,
is represented as
\begin{equation}
z=\frac{x^{2}}{8f_{0}}+ f - f_0  \qquad f_{0} = \frac{\sqrt{\left(L-f\right)^{2}+\frac{R_{0}^{2}}{2}}-\left(L-f\right)}{2}
\end{equation}
where $f$ is the focal length and $L$ and $R_{0}$ the positions
in $z$ and $x$ of any reference point on the parabola. An analogous
expression is used in the $z-y$ plane, orthogonal to the $z-x$ plane.

Unlike in the Lobster-Schmidt system, the Kirkpatrick-Baez is made of parabolic mirrors: this implies that the angular resolution on-axis can be much better than Lobster systems. Also the non-flat shape of the mirrors implies that, if the mirrors are close enough, on-axis stray light could be avoided; this would however require long enough mirrors, which would not be practical for a wide field telescope.

The Kirkpatrick-Baez design was optimized analogously to what is done in Section \ref{subsec:Angel_optimize} for Lobsters, finding the same results for the best optical parameters, including a mirror length of 40~mm.

\section{Comparison of the Optical Designs}\label{sec:Comparison-with-Other}

In this section we compare the performance achieved with the two Lobster configurations and with a Kirkpatrick-Baez configuration.

We also compare these ``new'' alternative designs with a standard Wolter-I configuration with the same focal length and aperture. For this, the geometrical design and equations are as in Ref. \citenum{2021hai4.book....3P}. Both the focal length and the length of the mirrors are the same as these found for the other designs above in Section \ref{subsec:Angel_optimize}. The mirror distance is chosen so that the next parabola is completely visible on-axis, but the mirror distance is imposed to be at least always greater than the minimum mirror separation for the other designs (so to have an analogous comparison). Baffles were simulated to have the entire parabola visible on-axis.
The Wolter-I has been designed so to enclose the same space occupancy, so made of a circle inscribed in the square area of the Lobsters and Kirkpatrick-Baez: this results in 138 nested shells from 35~mm to 138~mm radii. 

Other parameters would result in a different performance for the Wolter-I, but we are interested in comparing different optical designs with as much as possible the same parameters.
This design choice ensures a fair comparison of optical performance given identical satellite volume constraints, and is valid for a comparison based exclusively on the optical design. However, for an actual mission, one must also consider the weight, which depends on the mirror material, and so is outside the scope of this paper.

\begin{figure}
\begin{center}
\includegraphics[width=1\linewidth]{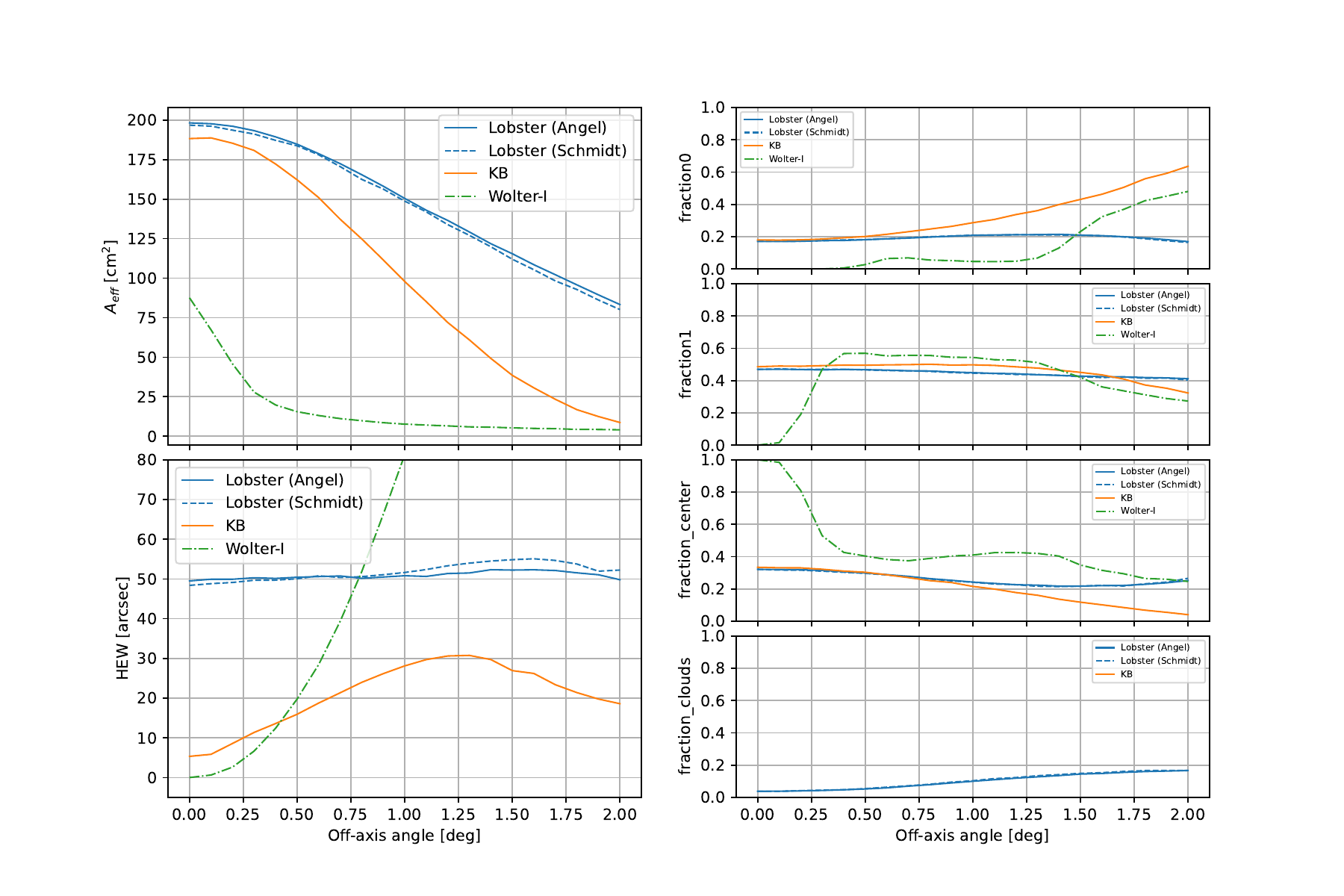}
\end{center}
\caption{\label{fig:Lobsters_comparison}
Effective area, half energy width and fractions of rays (see Figure \ref{fig:L_optimization} above for their explanation) comparison of different optical designs. The optical parameters are reported in Table \ref{tab:Optical-parameters-used}.
}
\end{figure}

Figure \ref{fig:Lobsters_comparison} shows a comparison of the performance of the different optical designs, we note that:
\begin{itemize}
\item The on-axis effective area is very similar for the Lobsters and Kirkpatrick-Baez, while it is smaller for the Wolter-I due to how mirrors are arranged (see discussion above). It then shows a very sharp decrease with the off-axis angle for the Wolter-I, and a less sharp decrease for the Kirkpatrick-Baez. The Lobster systems present a much better performance off-axis: this is because these systems are uniform in all directions, except for their limited physical size. This is the reason why, as seen in Figure \ref{fig:Lobster_max_size}, an increase of size is more effective in increasing the effective area off-axis than on-axis: the field of view can be increased just by building a larger system.
\item The ideal half energy width (so without considering the roughness or shape errors in the optics, outside the scope of this paper) is reached on-axis by Wolter-I systems and almost also by Kirkpatrick-Baez systems, but both these systems present a strong worsening going off-axis. For Lobster systems, the angular resolution depends only on the focal length and on the size of the capillaries (or mirrors separation) as $\propto \frac S f$, so it is constant over the entire field of view.
\item Lobster systems present an almost constant stray light and Kirkpatrick-Baez systems present a stray light increasing with the off-axis angle; the Wolter-I stray light is minimized due to the presence of baffles. The fraction of photons reflected on only one surface remains constant for the non-Wolter-I systems; crucially, the information contained within these single-reflection ``cross'' features is still valuable for source localization. We also note that the Kirkpatrick-Baez, because its mirrors are reflecting on only one side, does not present the  ``clouds'' of double reflected photons on the margin on opposite mirrors that are present for Lobsters. 
\end{itemize}

\begin{figure}
\begin{center}
\includegraphics[width=0.65\linewidth]{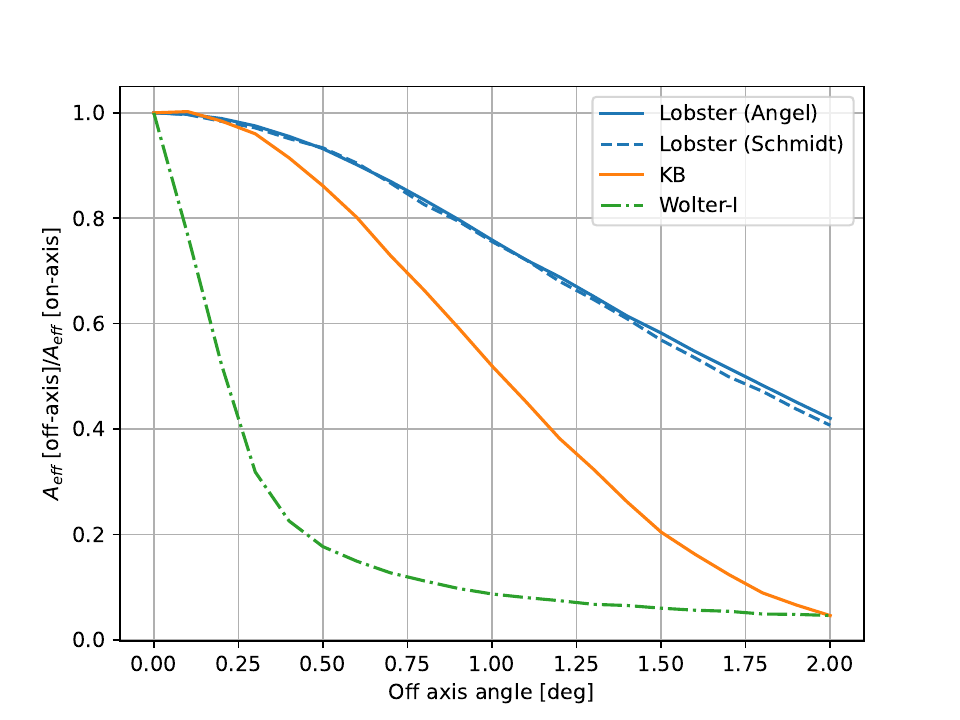}
\end{center}
\caption{\label{fig:Ratio-of-effective}
Ratio of effective area to half energy width, as a function of off-axis angle, for the various optical designs. 
The optical parameters are reported in Table \ref{tab:Optical-parameters-used}.
}
\end{figure}

The information on the effective area and on the half energy width can also be combined by taking their ratio, as shown in 
Figure \ref{fig:Ratio-of-effective}: on-axis all systems behave equally well, but the decrease off-axis is slower for Lobster systems.

Another method to compare the optics is to compute the integrated
product of effective area and field of view, also called \emph{etendue},
and defined as
\begin{equation}
E=\int\frac{A_{eff}\left(\theta\right)}{hew^{k}(\theta)}2\pi\theta\text{d}\theta\simeq\sum\frac{A_{eff}\left(\theta_{i}\right)}{hew^{k}(\theta)}\pi\left(\theta_{i,min}^{2}-\theta_{i,max}^{2}\right)\label{eq:etendue}
\end{equation}
the approximation in the second term of the above equation is how
we actually compute this quantity, using the values from Figure \ref{fig:Lobsters_comparison}.
We can give a different
weight to the half energy width through the power law index $k$ (to make this comparison realistic we sum in quadrature a 15~arcsec manufacturing error to the half energy width).
From Table \ref{tab:Etendue} we see that for a lower value of $k$,
Lobster Eye designs are favored, while for a higher value Kirkpatrick-Baez designs are favored.

\begin{table}[b]
\caption{\label{tab:Etendue}
Etendue as computed in equation \ref{eq:etendue}, where the parameter $k$ is the power law index of the half energy width. An error of 15~arcsec has been summed in quadrature to the half energy width to simulate manufacturing errors}
\begin{center}
\begin{tabular}{c|c|c|c|c}
$k$ & Angel & Schmidt & Kirkpatrick-Baez & Wolter-I\\
\hline 
0 & 1601 & 1568 & 808 & 98\\ \hline 
1 & 30 & 29 & 30 & 2.2\\ \hline 
2 & 0.6& 0.5 & 1.2 & 0.1 \end{tabular}
\end{center}
\end{table}

\section{Conclusion}\label{sec:Conclusion}
In this paper we compared different optical designs in X-rays, with the goal of building a telescope able to accurately follow up gravitational wave events in the region of sky constrained by gravitational wave interferometers, while fitting into a small mission size.

From Figure \ref{fig:Lobsters_comparison} we see that a $28\times28$~cm$^2$, 2500~mm focal length, Lobster design has an effective area $>$100~cm$^2$ up to 1.8~deg off-axis, so resulting in a 10~deg$^{2}$ field of view, while the half energy width remains at a steady 50~arcsec value over the entire field of view, thanks to the longer focal length with respect to existing Lobster Eyes. This would be ideal for a small mission dedicated to accurately localizing transients found as electromagnetic counterparts of gravitational waves. 
However stray light is not absent, so that further studies on the shape of the point spread function will have to be undertaken to better understand if its effect is negligible or not.
A Kirkpatrick-Baez design presents a faster decrease in performance off-axis, but much better angular resolution, so that a more advanced solution could be to employ an array of Kirkpatrick-Baez telescopes, each with its own detector. 

We compared these two designs with a Wolter-I system with analogous aperture and focal length: assuming no manufacturing errors, the Wolter-I system bring better angular resolution, and smaller stray light, but only in a very small field of view, and with smaller effective area, while over a larger field of view the Lobster Eye and Kirkpatrick-Baez designs provide a much better compromise.

We conclude that a single Lobster system, if big enough, is the simplest system to accommodate the scientific requirements for accurate localization of gravitational wave counterparts in X-rays.

\acknowledgments % equivalent to \section*{ACKNOWLEDGMENTS}       
 
This work was supported by fundamental research financing by the Italian National Institute of Astrophysics (INAF): MG-RSN5 1.05.24.07.05

% References
\bibliography{Comparing_optical_designs_citations} % bibliography data in report.bib

\begin{thebibliography}{10}

\bibitem{2023JCAP...07..068B}
{Branchesi}, M., {Maggiore}, M., {Alonso}, D., {Badger}, C., {Banerjee}, B.,
  {Beirnaert}, F., {Belgacem}, E., {Bhagwat}, S., {Boileau}, G., {Borhanian},
  S., {Brown}, D.~D., {Leong Chan}, M., {Cusin}, G., {Danilishin}, S.~L.,
  {Degallaix}, J., {De Luca}, V., {Dhani}, A., {Dietrich}, T., {Dupletsa}, U.,
  {Foffa}, S., {Franciolini}, G., {Freise}, A., {Gemme}, G., {Goncharov}, B.,
  {Ghosh}, A., {Gulminelli}, F., {Gupta}, I., {Kumar Gupta}, P., {Harms}, J.,
  {Hazra}, N., {Hild}, S., {Hinderer}, T., {Siong Heng}, I., {Iacovelli}, F.,
  {Janquart}, J., {Janssens}, K., {Jenkins}, A.~C., {Kalaghatgi}, C.,
  {Koroveshi}, X., {Li}, T. G.~F., {Li}, Y., {Loffredo}, E., {Maggio}, E.,
  {Mancarella}, M., {Mapelli}, M., {Martinovic}, K., {Maselli}, A., {Meyers},
  P., {Miller}, A.~L., {Mondal}, C., {Muttoni}, N., {Narola}, H., {Oertel}, M.,
  {Oganesyan}, G., {Pacilio}, C., {Palomba}, C., {Pani}, P., {Pasqualetti}, A.,
  {Perego}, A., {P{\'e}rigois}, C., {Pieroni}, M., {Piccinni}, O.~J.,
  {Puecher}, A., {Puppo}, P., {Ricciardone}, A., {Riotto}, A., {Ronchini}, S.,
  {Sakellariadou}, M., {Samajdar}, A., {Santoliquido}, F., {Sathyaprakash},
  B.~S., {Steinlechner}, J., {Steinlechner}, S., {Utina}, A., {Van Den Broeck},
  C., and {Zhang}, T., ``{Science with the Einstein Telescope: a comparison of
  different designs},'' {\em \jcap}~{\bf 2023},  068 (July 2023).

\bibitem{2010MNRAS.405..877C}
{Conconi}, P., {Campana}, S., {Tagliaferri}, G., {Pareschi}, G., {Citterio},
  O., {Cotroneo}, V., {Proserpio}, L., and {Civitani}, M., ``{A wide field
  X-ray telescope for astronomical survey purposes: from theory to practice},''
  {\em \mnras}~{\bf 405},  877--886 (June 2010).

\bibitem{1979ApJ...233..364A}
{Angel}, J.~R.~P., ``{Lobster eyes as X-ray telescopes.},'' {\em \apj}~{\bf
  233},  364--373 (Oct. 1979).

\bibitem{1975NucIM.127..285S}
{Schmidt}, W.~K.~H., ``{A proposed X-ray focusing device with wide field of
  view for use in X-ray astronomy},'' {\em Nuclear Instruments and
  Methods}~{\bf 127},  285--292 (Jan. 1975).

\bibitem{1948JOSA...38..766K}
{Kirkpatrick}, P. and {Baez}, A.~V., ``{Formation of optical images by
  x-rays},'' {\em Journal of the Optical Society of America (1917-1983)}~{\bf
  38},  766 (Sept. 1948).

\bibitem{2021hai4.book...85W}
{Willingale}, R., ``{Lobster Eye Optics},'' in [{\em The WSPC Handbook of
  Astronomical Instrumentation, Volume 4: X-Ray Astronomical
  Instrumentation,}{\nolinebreak\hspace{0.1em}]},  {Burrows}, D.~N., ed.,
  85--106 (2021).

\bibitem{2022ApJ...941L...2Z}
{Zhang}, C., {Ling}, Z.~X., {Sun}, X.~J., {Sun}, S.~L., {Liu}, Y., {Li}, Z.~D.,
  {Xue}, Y.~L., {Chen}, Y.~F., {Dai}, Y.~F., {Jia}, Z.~Q., {Liu}, H.~Y.,
  {Zhang}, X.~F., {Zhang}, Y.~H., {Zhang}, S.~N., {Chen}, F.~S., {Cheng},
  Z.~W., {Fu}, W., {Han}, Y.~X., {Li}, H., {Li}, J.~F., {Li}, Y., {Liu}, P.~R.,
  {Ma}, X.~H., {Tang}, Y.~J., {Wang}, C.~B., {Xie}, R.~J., {Yan}, A.~L.,
  {Zhang}, Q., {Jiang}, B.~W., {Jin}, G., {Li}, L.~H., {Qiu}, X.~B., {Su},
  D.~T., {Sun}, J.~N., {Xu}, Z., {Zhang}, S.~K., {Zhang}, Z., {Zhang}, N.,
  {Bi}, X.~Z., {Cai}, Z.~M., {He}, J.~W., {Liu}, H.~Q., {Zhu}, X.~C., {Cheng},
  H.~Q., {Cui}, C.~Z., {Fan}, D.~W., {Hu}, H.~B., {Huang}, M.~H., {Jin}, C.~C.,
  {Li}, D.~Y., {Pan}, H.~W., {Wang}, W.~X., {Xu}, Y.~F., {Yang}, X., {Zhang},
  B., {Zhang}, M., {Zhang}, W.~D., {Zhao}, D.~H., {Bai}, M., {Ji}, Z., {Liu},
  Y.~R., {Ma}, F.~L., {Su}, J., {Tong}, J.~Z., {Wang}, Y.~S., {Zhao}, Z.~J.,
  {Feldman}, C., {O'Brien}, P., {Osborne}, J.~P., {Willingale}, R., {Burwitz},
  V., {Hartner}, G., {Langmeier}, A., {M{\"u}ller}, T., {Rukdee}, S.,
  {Schmidt}, T., {Kuulkers}, E., and {Yuan}, W., ``{First Wide Field-of-view
  X-Ray Observations by a Lobster-eye Focusing Telescope in Orbit},'' {\em
  \apjl}~{\bf 941},  L2 (Dec. 2022).

\bibitem{2022hxga.book...86Y}
{Yuan}, W., {Zhang}, C., {Chen}, Y., and {Ling}, Z., ``{The Einstein Probe
  Mission},'' in [{\em Handbook of X-ray and Gamma-ray
  Astrophysics}{\nolinebreak\hspace{0.1em}]},   86 (2022).

\bibitem{2023ExA....55..487G}
{G{\"o}tz}, D., {Boutelier}, M., {Burwitz}, V., {Chipaux}, R., {Cordier}, B.,
  {Feldman}, C., {Ferrando}, P., {Fort}, A., {Gonzalez}, F., {Gros}, A.,
  {Hussein}, S., {Le Duigou}, J.~M., {Meidinger}, N., {Mercier}, K., {Meuris},
  A., {Pearson}, J., {Renault-Tinacci}, N., {Robinet}, F., {Schneider}, B., and
  {Willingale}, R., ``{The scientific performance of the microchannel X-ray
  telescope on board the SVOM mission},'' {\em Experimental Astronomy}~{\bf
  55},  487--519 (Apr. 2023).

\bibitem{2021hai4.book....3P}
{Pareschi}, G., {Spiga}, D., and {Pelliciari}, C., ``{X-ray Telescopes Based on
  Wolter-I Optics},'' in [{\em The WSPC Handbook of Astronomical
  Instrumentation, Volume 4: X-Ray Astronomical
  Instrumentation,}{\nolinebreak\hspace{0.1em}]},   3--31 (2021).

\end{thebibliography}
\bibliographystyle{spiebib} % makes bibtex use spiebib.bst

\end{document}